\documentclass[aps,prc,twocolumn,preprintnumbers,amsmath,amssymb]{revtex4}
\usepackage{graphicx}% Include figure files
\usepackage{dcolumn}% Align table columns on decimal point
\usepackage{bm}% bold math
\usepackage{color}%
\usepackage{epstopdf}
\usepackage{dcolumn}% Align table columns on decimal point
\usepackage[sort&compress]{natbib}
\usepackage{float}
\usepackage[utf8]{inputenc}
\usepackage[T1]{fontenc}
\begin{document}
\title{Inclusive $\alpha$ Production for $^{6}$Li+$^{51}$V System\\}
\author{\large C. Joshi$^{1}$}
\author{\large H. Kumawat$^{2,3}$\footnote{harphool@barc.gov.in}}
\author{\large V. V. Parkar$^{2,3}$}
\author{\large D. Dutta$^{2,3}$}
\author{\large S. V. Suryanarayana$^{2}$}
\author{\large V. Jha$^{2,3}$}
\author{\large R. K. Singh$^1$}
\author{\large N.L . Singh$^1$}
\author{S. Kailas$^{2,3,4}$}%
\affiliation{$^1$Department of Physics,The M. S. University of Baroda, Vadodara 390002, India}
\affiliation{$^2$Nuclear Physics Division, Bhabha Atomic Research Centre, Mumbai 400085, India}
\affiliation{$^3$Homi Bhabha National Institute, Anushaktinagar, Mumbai 400094, India}
\affiliation{$^4$UM-DAE Centre for Excellence in Basic Science, Mumbai 400098, India}
\date{\today}

\begin{abstract}
 \noindent\textbf{Background:} Experimental and theoretical studies of nuclear reaction mechanism to understand relative contribution in large $\alpha$ production due to breakup, incomplete fusion and transfer reactions induced by weakly bound projectiles near Coulomb barrier are important.\\ 
 \noindent\textbf{Purpose:} Measurement of angular distributions and energy spectra of $\alpha$ and deuterons through breakup, transfer and incomplete fusion processes to dis-entangle their relative contributions and to investigate relative importance of breakup-fusion compared to transfer.\\ 
 \noindent\textbf{Methods:} Inclusive $\alpha$ production cross-sections have been measured for $^6$Li + $^{51}$V system near Coulomb barrier energies. Theoretical calculations for estimation of various reaction channels contributing to $\alpha$ production have been performed with finite range coupled reaction method using \textsc{FRESCO} code. The cross-sections from non-capture breakup (NCBU) ($\alpha$ + \textit{d}) and 1\textit{n}, 1\textit{p}, and 1\textit{d} transfer channels, compound nuclear decay channel and incomplete fusion (ICF) leading to $\alpha$ production were estimated to get the cumulative production cross-sections.\\ 
 \noindent\textbf{Results:} Contributions from breakup, transfer and incomplete fusion channels could reproduce the integral direct $\alpha$ production cross-sections and their angular distributions quite well. The direct $\alpha$ production cross-sections are in agreement with other targets. The $\alpha$ production cross-sections are higher compared to the deuteron production.\\ 
 \noindent\textbf{Conclusions:} Kinematic analysis of the energy spectra of $\alpha$ particles and deuterons suggest that $\alpha$ particle spectra is dominated by breakp-fusion and deuteron spectra have contribution of breakup and transfer reactions. A systematic study of direct $\alpha$ production with various targets follow a universal behavior on average  but noticeable differences are observed for different targets. A ratio of $\alpha$ and deuteron yields for a wide mass range of targets shows a saturation above barrier and an increasing production of $\alpha$ particles relative to deuteron around Coulomb barrier.
\end{abstract}

\maketitle
\section{\label{sec:level1}Introduction}
The transfer and breakup reactions induced by weakly bound projectiles are important reaction channels around Coulomb barrier. The breakup of weakly bound light nuclei such as $^6$Li with cluster structure ($\alpha$ + \textit{d}) is a well established phenomenon while moving in the proximity to the field of target nucleus \cite{lei17,pakou03,kelly2000,souza09,kumawat10}. Numerous observations exist for larger yield of $\alpha$ particles in comparison to its complementary constituent and the mechanism lying behind is still the current interest of investigation for projectiles with cluster structures like $^{6,8}$He, $^{6,7}$Li, $^{7,9}$Be  \cite{canto2015,sgouros16,lei17,kolata16}. Large yield of inclusive $\alpha$ particles compared to complimentary deuteron cluster implies existence of several processes apart from breakup \cite{PhysRevC.67.044607, HUGI1981173, PhysRevC.99.024620, PFEIFFER1973545,lei17,souza09}. A systematic understanding of inclusive $\alpha$ production and the different reaction channels contributing to it, have not been yet clearly identified by inclusive and exclusive measurement, and theoretical estimations \cite{PhysRevC.85.014612, santra09, PhysRevC.88.064603, PhysRevC.83.034616}. Three types of measurements are performed to understand reaction mechanism with weakly bound nuclei; a) elastic scattering: to get reaction cross-section and understand potential behavior near Coulomb barrier, effect of breakup and transfer on it. b) exclusive measurement: particle-particle or particle-$\gamma$ coincidence to get the cross-sections of the major channels. c) inclusive measurement: to get contribution from all major and minor channels.

A large number of studies were performed with weakly bound projectile $^6$Li to have better understanding of breakup influences on elastic scattering and fusion \cite{lei17,canto2015,sgouros16,kumawat20,kumawat12}. The yield of $\alpha$ particles is significantly higher than that of deuterons which indicates deuteron transfer and breakup-fusion is favored over $\alpha$, and there are many more reaction channels that produce $\alpha$ particles as compared to deuterons. Inclusive $\alpha$ production incorporates distinct reaction mechanisms, right from breakup to compound nuclear evaporation along with nucleon transfer trailed by breakup; incomplete fusion or transfer of a cluster. 
The complete fusion (CF) is found to be suppressed by $\approx$ 30\% for mid to heavy mass targets \cite{kumawat12} but less suppression was reported for light mass targets \cite{PhysRevC.94.044605, PhysRevC.91.044619, Sinha2017,PhysRevC.5.1835,PhysRevC.90.024615}. The suppression is due to breakup, transfer and incomplete fusion processes although, there is a difficulty in separation of CF and ICF channels for light mass targets and the conclusions are model dependent. As $\alpha$ production is the main channel in all direct reactions (transfer and ICF) other than breakup, it is interesting to understand if direct $\alpha$ production also gets suppressed in light mass region like $^{51}$V. 

The $\alpha$ energy spectrum from an exclusive measurement for $^7$Li+ $^{93}$Nb system \cite{PANDIT2021136570} suggested that transfer is dominant while other studies have suggested a mechanism of breakup-fusion to be more important \cite{PhysRevLett.98.152701}. In the case of transfer, the \textit{Q} value is shared and contributes to high energy $\alpha$ production for positive \textit{Q} value reactions but in the case of breakup-fusion the $\alpha$ particles do not get extra energy from the positive \textit{Q} value. Theoretical studies could explain $\alpha$ production from cluster transfer mechanism for $^7$Li case \cite{lei17,PhysRevLett.122.042503}. In the case of $^6$Li projectile neutron transfer was suggested to be responsible for $\approx$ 50\% direct $\alpha$ production \cite{HUGI1981173}, transfer was assumed to be responsible without distinguishing different transfer channels \cite{PhysRevC.76.054601}. Presently, none of the coupled channel codes are competent enough to include breakup and transfer in a comprehensive calculation. Several reports \cite{santra09,kumawat10,acosta11,pandit17} manifest this aspect to define breakup cross section and have done exclusive and inclusive measurements to find the solution of this open question of various contributions to inclusive $\alpha$ production.

In the present work, the energy and angular distributions of inclusive $\alpha$, deuteron, and the integral cross sections for the $^6$Li+$^{51}$V system are reported. Kinematic analysis of $\alpha$  and deuteron energy spectra was performed to understand dominant process between transfer and breakup-fusion. Theoretical calculations for breakup and various transfer channels are performed to interpret the experimental data. The results manifest contribution from breakup and transfer channels. Systematics of direct $\alpha$ cross-section and its relation with deuteron cross-sections were performed. The article contains following outline. Sec. II is dedicated to experimental details. Sec. III describes the data reduction procedure and brief discussion.  Kinematic disentanglement for origin of $\alpha$ particles and deuterons by different processes is discussed in sec. IV. Theoretical analysis using statistical model, CDCC, coupled reaction calculation of 1\textit{n}, 1\textit{p} and 1\textit{d}  transfer using FRESCO are described in Sec. V. Systematic study of direct $\alpha$  production cross-section, ratio of $\alpha$ and deuteron cross-sections are described in sec. VI. Summary is given in sec. VII.

\section{\label{sec:level2}Experimental details}
\begin{figure}
\includegraphics[scale=0.4]{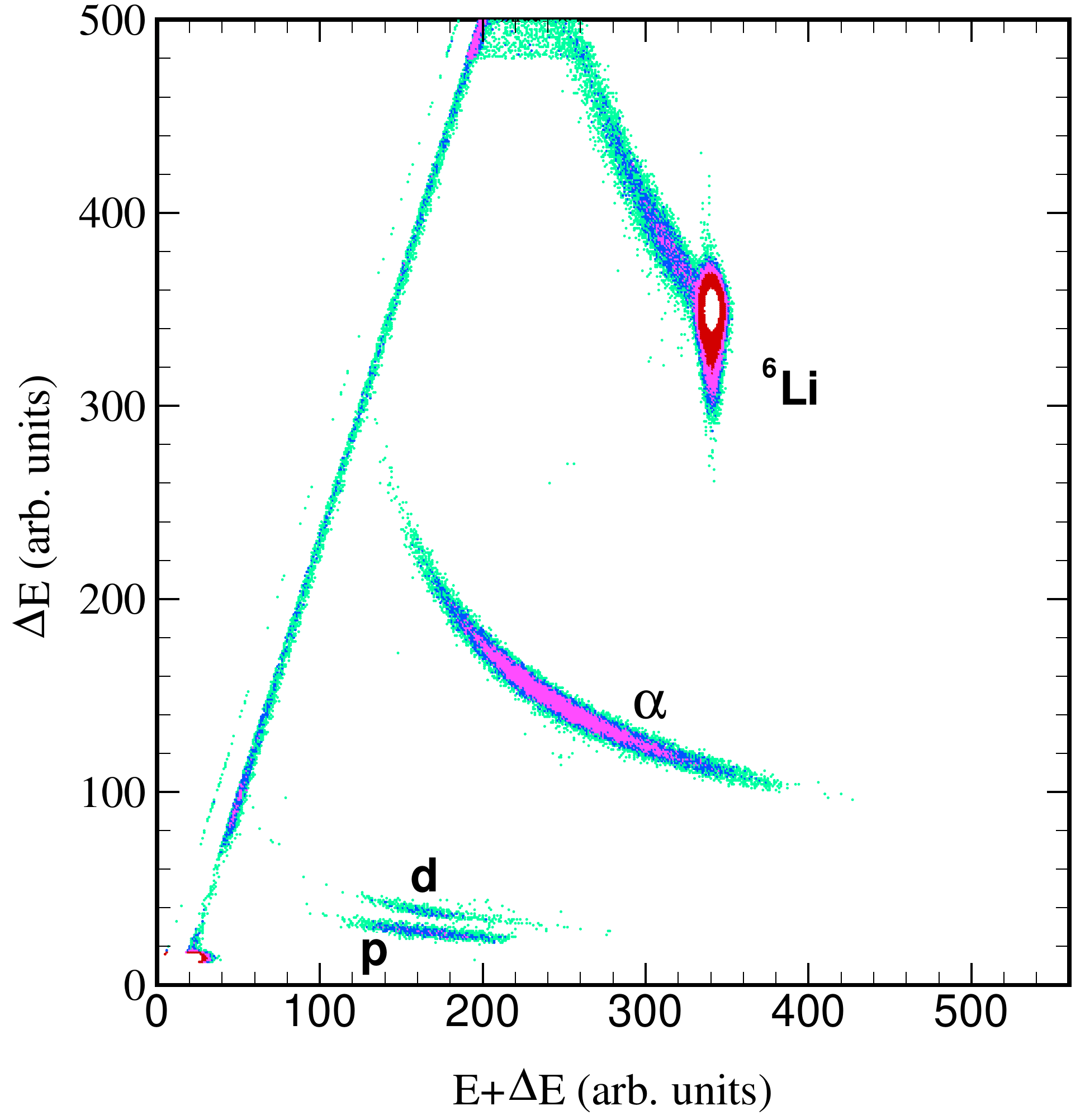}
\caption{\label{fig:spec} The typical bi-parametric $\Delta$\textit{E} - \textit{E}$_{Total}$ plot for the $^6$Li+$^{51}$V system at \textit{E}$_{lab}$ = 19.7 MeV, $\theta_{lab}$ = 30$^\circ$.}
\end{figure}

Details of the experimental setup are given in our earlier publication \cite{kumawat20} where breakup threshold anomaly was reported and only a short summary is given here for completeness. The experiment was performed at 14-UD BARC-TIFR Pelletron-Linac accelerator facility, Mumbai, India with $^6$Li$^{3+}$ beam at energies 14, 20, 23 and 26 MeV. The beam current was ranging between 5-28 nA. The beam was incident on a self-supported $^{51}$V target of thickness 1.17 mg/cm$^2$. Beam energies were corrected for the energy loss in the target (13.6, 19.7, 22.7 and 25.7 MeV). The detection system was consisting of a set of four solid state silicon surface barrier telescope detectors in $\Delta$\textit{E} + \textit{E} arrangement and two monitors at $\pm$10$^\circ$ for absolute normalization. The angles covered by telescope detectors were 14$^\circ$ to 170$^\circ$ in lab. frame. A typical measured $\Delta$\textit{E} - \textit{E}$_{Total}$ 2D-plot at \textit{E}$_{lab}$ = 19.7 MeV and $\theta_{lab}$ = 30$^\circ$ is given in Fig. \ref{fig:spec}. The statistical errors were $\approx$ 1\% at forward angles which gradually increase up to $\approx$ 10\% above $\theta_{lab}$ = 70$^\circ$ for 19.7, 22.7 and 25.7 MeV energies. In the case of 13.6 MeV energy, the statistical errors were less than 5\% at all angles. The data were recorded using the Linux based data acquisition system, LAMPS \cite{lamps}.

\section{\label{sec:level3} Experimental energy and angular distributions}

\begin{figure}
\includegraphics[scale=0.40]{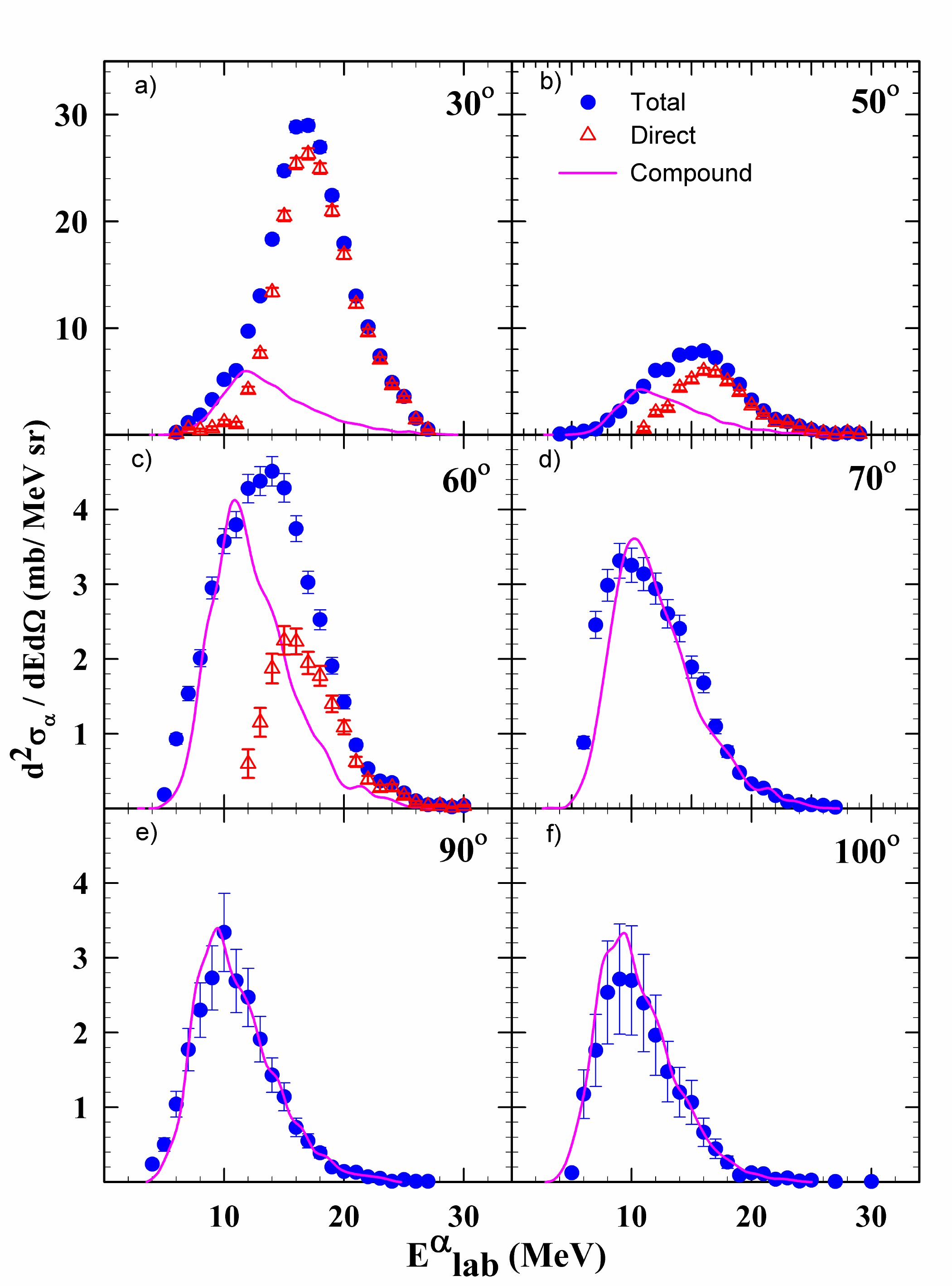}
\caption{\label{fig:enespec26}Energy spectra of $\alpha$ particles at E$_{lab}$ = 25.7 MeV and for various lab. angles for $^{6}$Li + $^{51}$V system. The experimental data for total and direct $\alpha$ production are presented by filled circles and hollow triangles, respectively. The compound nuclear contribution from \textsc{PACE} is represented by solid line.}
\end{figure}
\begin{figure}
\includegraphics[scale=0.4]{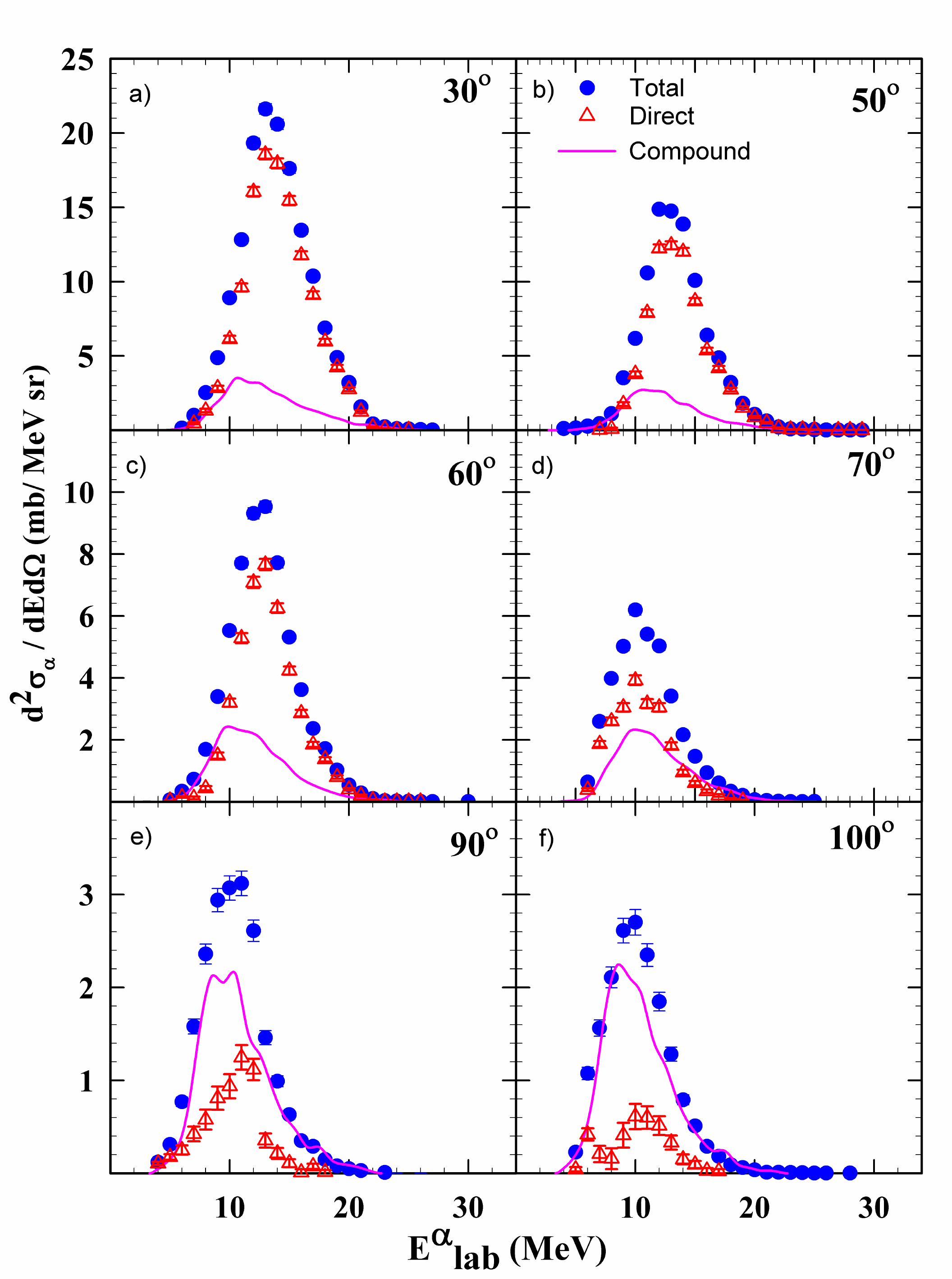}
\caption{\label{fig:enespec20}Same as given in Fig. \ref{fig:enespec26}. but at E$_{lab}$ =  19.7 MeV.}
\end{figure}
\begin{figure}
\includegraphics[scale=0.45]{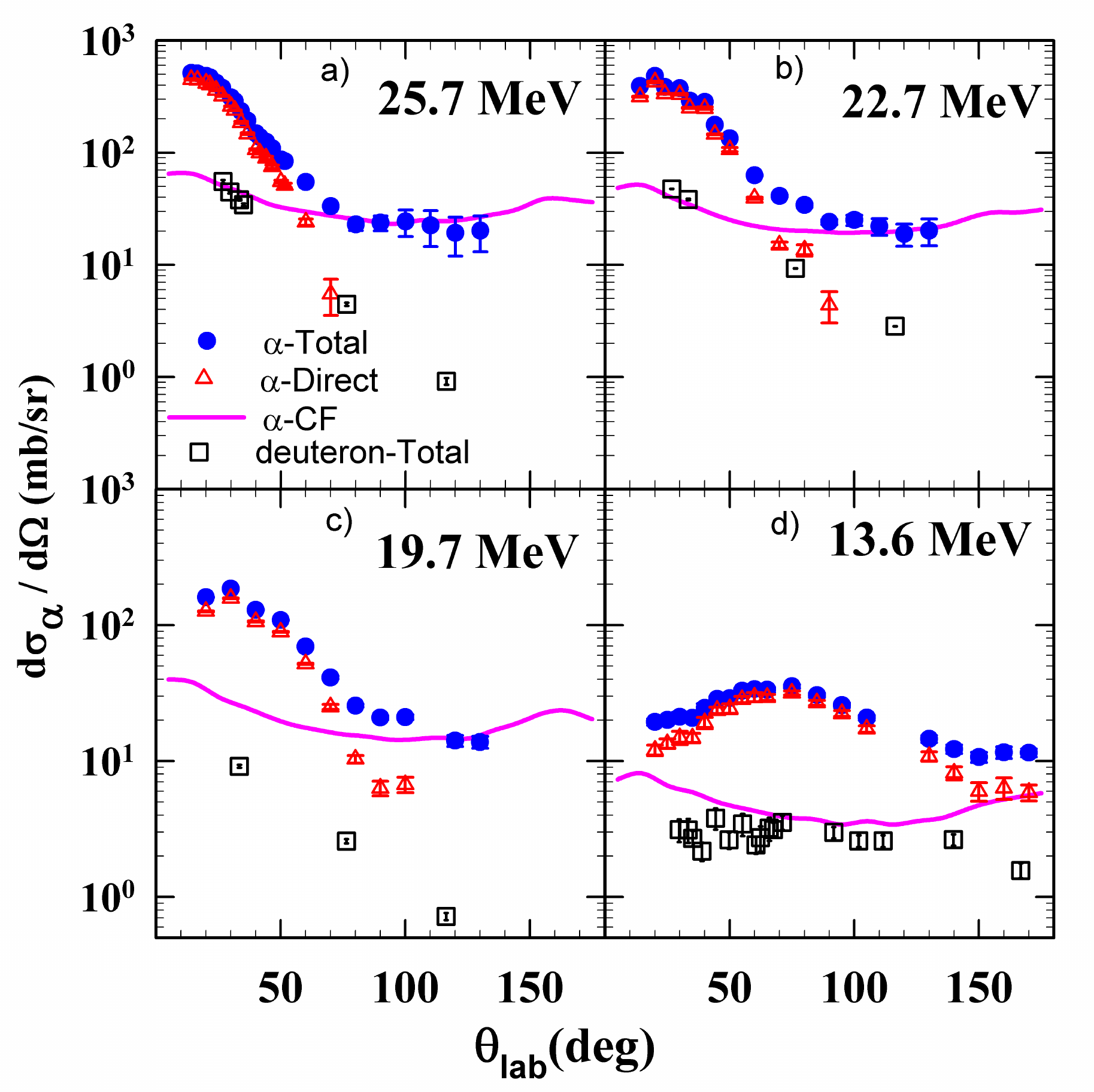}
\caption{Energy integrated angular distribution of $\alpha$ particles at different energies a) 25.7 MeV, b) 22.7 MeV, c) 19.7 MeV and d) 13.6 MeV for $^{6}$Li + $^{51}$V system. The solid circles represent total $\alpha$  cross section, solid line is contribution from compound nuclear reaction, and direct or non-compound (total - compound) are depicted by hollow triangles. Total deuteron cross-sections are shown by hollow squares.}
\label{fig:inclspec}
\end{figure}

The energy spectra of $\alpha$ particles at various angles are shown in Fig. \ref{fig:enespec26} and Fig. \ref{fig:enespec20} for 25.7 MeV and 19.7 MeV, respectively. The experimental energy spectra include $\alpha$ particles originating from direct and compound nuclear reactions. The calculated values from compound nuclear reaction have dominant contribution well above grazing angles while the direct contributions peaks around grazing angles. The grazing angles for 25.7, 22.7, 19.7 and 13.6 MeV are $\approx$ 31$^\circ$, 36$^\circ$, 45$^\circ$, and 91$^\circ$,  respectively which were obtained from elastic scattering data \cite{kumawat20}. The direct contribution is deduced by subtracting the calculated compound nuclear contribution from measured $\alpha$ production. The experimental energy and angular distributions are matching well with compound nuclear contributions at higher angles (70$^\circ$ and above for 25.7 MeV and above 100$^\circ$ for 19.7 MeV) to offer the only contribution from complete fusion process above these angles. Direct $\alpha$ is dominated around grazing angles. The energy integrated $\alpha$ particle yields were obtained at different angles. The energy integrated measured differential angular cross sections were obtained \cite{kumawat10} using the following equation
\begin{equation}
\frac{d\sigma_\alpha}{d\Omega} = \frac{Y_\alpha}{Y_{el}}\times\frac{d\sigma_{el}}{d\Omega}
\end{equation}
Here, \textit{Y}$_\alpha$, \textit{Y}$_{el}$ are $\alpha$ particle and elastic scattering yields, $d\sigma_{el}$/$d\Omega$ is the elastic scattering cross section, as reported in Ref. \cite{kumawat20}. Angular distributions of $\alpha$ particle production cross sections are shown in Fig. \ref{fig:inclspec}. It is clear that the cross-sections well above grazing angles are dominated by evaporation through the compound nuclear or complete fusion reaction whereas breakup and transfer following $\alpha$ particles are peaking near grazing angles.

The angle integrated direct $\alpha$ cross sections at each energy were obtained by fitting the Gaussian shape to (d$\sigma$/d$\Omega$) $\times$ 2$\pi$sin$\theta$ distribution and an integral $\alpha$ cross section was deduced using equation
\begin{equation}
\sigma_{\alpha}= \int_{0}^{2\pi} d\phi \int_{0}^{\pi} {\frac{d\sigma_\alpha(\theta)}{d\Omega}} sin\theta d\theta
\end{equation}
The deduced experimental direct $\alpha$ cross sections along with errors are given in Table \ref{tab:incl}. The errors were obtained due to fitting errors in the three parameters (strength, mean, and width) of the Gaussian distributions. Maximum and Minimum cross-sections were obtained by adding these errors to the mean values of the parameters, and thus errors were deduced in the cross-sections. The angular range of measurements was very well covered around grazing angles for 19.7 and 13.6 MeV energies hence, the errors in the data were less. The angular distributions for 22.7 and 25.7 MeV energies were very much forward peaked and the measurements were done from 14$^\circ$ onward, hence, the deduced errors were relatively higher at these energies.
\begin{center}
\begin{table*} [htbp]
\caption {\label{tab:incl} Experimental direct $\alpha$ production
($\sigma^{direct}_\alpha$) and total deuteron production cross-sections ($\sigma^{total}_d$) deduced from integral of measured angular distributions, calculated cross sections for $\alpha$ production by compound nuclear reactions $\sigma^{CN}_{\alpha}$, non-capture breakup ($\sigma_\alpha^{NCBU}$), 1\textit{n}transfer ($\sigma_{1-n}^{TC}$), 1p-transfer ($\sigma_{1-p}^{TC}$), 1\textit{d}transfer ($\sigma_{1-d}^{TC}$), ICF ($\sigma_{\alpha}^{ICF}$) and total calculated direct $\alpha$ production ($\sigma^{cal.}_{\alpha}$). Transfer calculations are performed using \textsc{FRESCO} and compound nuclear calculations are done with \textsc{PACE} code. The complete fusion cross-sections given by \textsc{PACE} are denoted by $\sigma^{CF}_{pace}$.}
\begin{ruledtabular}
\begin{tabular}{cccccccccccc}
\centering
$E_{lab}$ & $\sigma^{direct}_\alpha$ &$\sigma_{d}$ & $\sigma^{CN}_{\alpha}$  &$\sigma^{CF}_{pace}$ 
&$\sigma_\alpha^{NCBU}$ 
&$\sigma_{1-n}^{TC}$ 
&$\sigma_{1-p}^{TC}$ 
&$\sigma_{1-d}^{TC}$ 
&$\sigma_{\alpha}^{ICF}$ 
&$\sigma^{cal.}_{\alpha}$\\
MeV &(mb)& (mb)&(mb) &(mb)  &(mb) &(mb) &(mb) &(mb) &(mb)  &(mb)\\
\hline
25.7 & 510 $\pm$ 46 &136 $\pm$ 27 & 387 & 1056 & 62  & 43 & 16 & 61 & 277 & 459 \\
22.7 & 490 $\pm$ 68& 144 $\pm$ 38 & 304 & 963  & 58  & 36 & 16 & 61  &253 & 424 \\
19.7 & 403 $\pm$ 11&  & 230 & 826  &  52  & 25 & 15 & 56 &  217& 365  \\
13.6 & 239 $\pm$ 8 & 36 $\pm$ 7 & 54  & 279  & 25  & 24 & 10  & 42 & 73 & 174 \\
\end{tabular}
\end{ruledtabular}
\end{table*}    
\end{center}
\begin{figure}[h]
\includegraphics[scale=0.40]{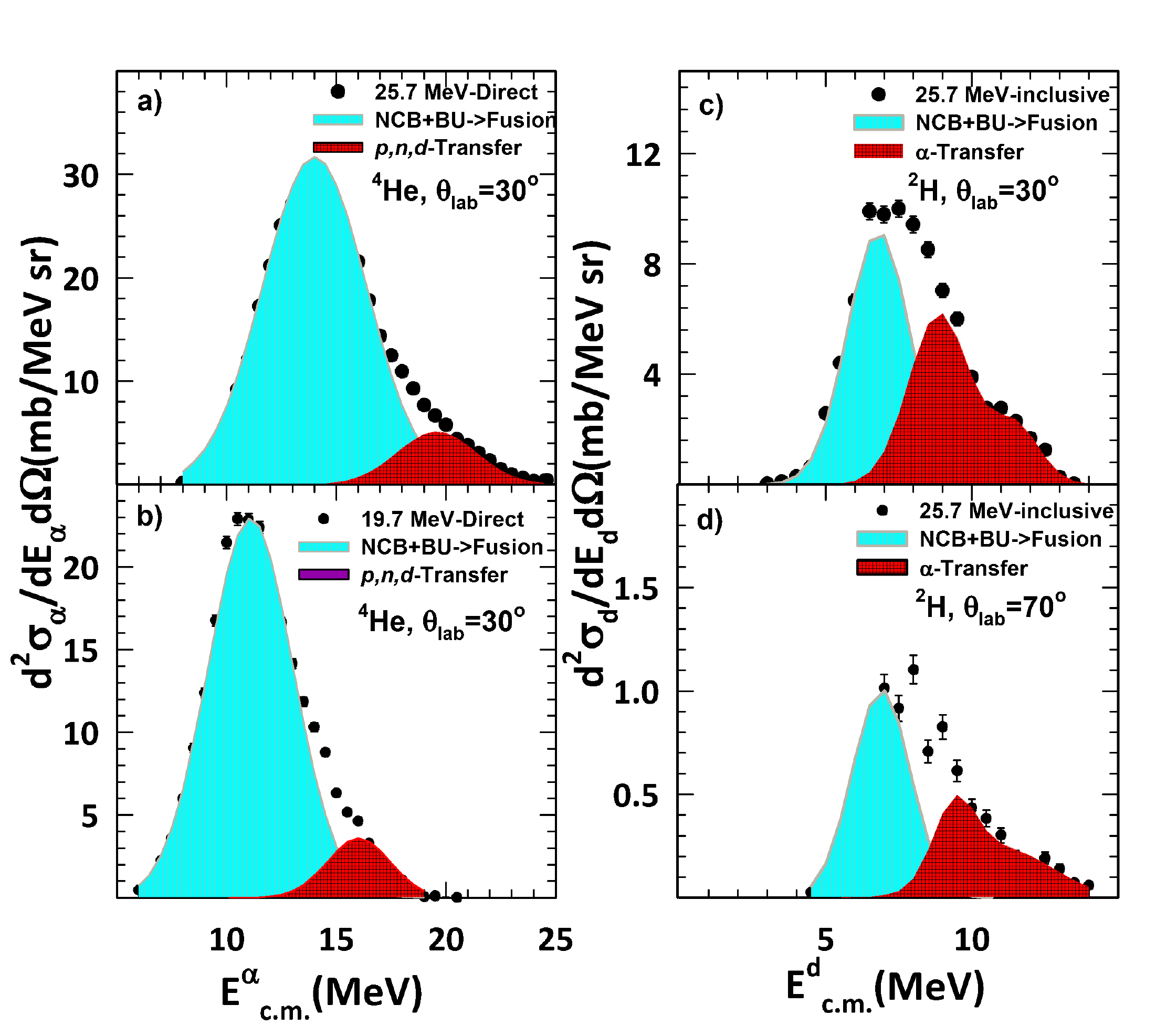}
\caption{\label{fig:disent}$\alpha$-energy spectra at $\theta_{lab}$=30$^\circ$ for a) 25.7 and b) 19.7 MeV. Deutron spectra are given at c) $\theta_{lab}$=30$^\circ$ and d) 70$^\circ$ degrees for 25.7 MeV beam energy. Non-capture breakup (NCBU), breakup-fusion are shown by shaded region (cyan) and \textit{n, p, d} and $\alpha$ transfers are shown by red shaded area (see section \ref{sec:level4} for details).}
\end{figure}
\section{\label{sec:level4} Kinematic disentanglement of $\alpha$ particle energy spectra}
Detailed analysis of the energy spectra was carried out and energy centroids were calculated using kinematics for different processes and are given in Table \ref{tab:qopt}. The elastic or non-capture breakup contribution is in general very well reproduced by CDCC calculations using \textsc{FRESCO} and is $\approx$ 10 \% of direct $\alpha$ cross-section as given in Table \ref{tab:incl}. In the case of transfer reactions, if fusion followed by breakup takes place then energy from \textit{Q} value is not shared with the out going fragments like $\alpha$, $^{5}$He and $^{5}$Li for deuteron, proton and neutron transfer, respectively. The centroid energies in Table \ref{tab:qopt} are given for the cluster transfer where the energy from \textit{Q} value is not shared with the outgoing fragment. Neutron transfer has only one value as optimum \textit{Q} value ($Q_{opt}$) is zero. In the case of sharing of \textit{Q} value with the outgoing $\alpha$ particles, following possibilities are there for the present study at 25.7 MeV beam energy

(a) deuteron transfer: the $\alpha$ particles from ground state transfer should peak around 25 MeV; for transfer to lower states of the target but sufficient to emit one neutron, should peak at $\approx$ 22 MeV and transfer to higher states of the target but sufficient to emit two neutrons, should peak at $\approx$ 19 MeV. 
(b) neutron transfer: the $\alpha$ particles from ground state transfer should peak around $\approx$ 21 MeV. 
(c) proton transfer: the $\alpha$ particles from ground state transfer should peak around $\approx$ 17 MeV. The calculations (Table \ref{tab:incl}) suggest this process to be very small compared to other two processes. 

In all three cases of the transfer reactions if \textit{Q} value is shared with the outgoing fragments then there should be a peak around 20 MeV. The energy spectra were fitted with two Gaussians with all free parameters.The contributions were estimated using areas under Gaussian peaks as shown in Fig. \ref{fig:disent}. The relative peak area around 20 MeV was estimated to be $\approx$ 10 \% which corresponds to  deuteron, proton and neutron transfer, altogether. Other contributions to $\alpha$ production are from breakup fusion and NCBU which are under the major peak. The NCBU is $\approx$ 10 \% of direct $\alpha$ cross-section as shown in Table \ref{tab:incl} and the remaining $\alpha$ particles seem to originate from breakup fusion where \textit{Q} value does not boost the $\alpha$ particle energy. Small contribution of such boosted $\alpha$ particles perhaps indicates that breakup-fusion is the dominant process for $^6$Li projectile compared to transfer while the transfer was reported to be dominant in the case of $^7$Li+$^{93}$Nb \cite{PANDIT2021136570}.

An estimate of the $\alpha$ particles from breakup fusion or incomplete fusion has been made as suggested by Jha et al. \cite{JHA20201} and mentioned here. As the complete fusion is suppressed by $\approx$ 35\% for $^6$Li induced reactions and 75\% of this is originated from deuteron capture reaction producing ICF $\alpha$. The cross-sections for ICF $\alpha$ particles were estimated using complete fusion cross-sections given in Table \ref{tab:incl}. The ICF $\alpha$ and other direct $\alpha$ cross-sections estimated from different reaction channels are given in Section \ref{sec:level5} and are close to the experimental values as shown in Fig. \ref{fig:inclusive}. It was not possible to separate ICF $\alpha$ and breakup $\alpha$ particles as both peaks are in the same region. In the case of deuteron energy spectra, it has mixture of breakup and $\alpha$ transfer. The breakup cross-sections and integral deuteron production suggest that both processes have almost equal share of it.

\begin{center}
\begin{table}
\caption{\label{tab:qopt}Kinematic parameters for $\alpha$ transfer channel for $^{6}$Li + $^{51}$V system. $Q_{gg}$ represent transfer to ground state, optimum \textit{Q} value ($Q_{opt}$) is calculated at leading order according to Ref. \cite{PhysRevC.17.2245}, centroid energy of the direct-$\alpha$ for the respective channels in the c.m. system ($\overline{E}_\alpha$=$E_{c.m.}$+$Q_{opt}$). $^{5}$Li$\rightarrow$ $\alpha + p$ and $^{5}$He $\rightarrow$ $\alpha + n$ are assumed to be broken giving $\alpha$ particles. The energies of breakup constituents are calculated as per mass ratio.}
\begin{ruledtabular}
\begin{tabular}{c c c c c c c}
 $E_{c.m.}$ & &\textit{d} trans. & \textit{p} trans. & \textit{n} trans. & breakup  \\
  (MeV) & & (MeV) & (MeV)  & (MeV) & (MeV) \\
\hline
  23.0 &Q$_{gg}$                 & 14.7  & 6.1   & 1.65   & -1.47 \\
      & Q$_{opt}$                & -7.1  &  -7.1 &  0.0   & \\
      & $\overline{E}_\alpha$    & 16.0  & 13.4  & 21.2   &14.3\\
20.3  & Q$_{opt}$                & -6.3  & -6.3  &  0.0   & \\
      & $\overline{E}_\alpha$    & 14.1  & 11.9  & 19.1   &12.6\\
17.6  & Q$_{opt}$                & -5.5  & -5.5  &  0.0   & \\
      & $\overline{E}_\alpha$    & 12.3  & 10.4  & 16.9   &10.8\\
\end{tabular}
\end{ruledtabular}
\end{table}
\end{center}

\section{\label{sec:level5}Theoretical Analysis }

\subsection{\label{sec:level41}Statistical Model Calculations}
The contribution from compound nuclear reaction for $\alpha$ production 
cross section was calculated using statistical model PACE \cite{gavron}. The angular distributions and energy spectra show resemblance well above grazing angles with experimental data. The Ignatyuk  prescription \cite{ignatyuk} of level density with parameter ($\tilde{a}$ = \textit{A}/10 MeV$^{-1}$, \textit{A} = mass number) was used. The optical potentials used in PACE are from C. M. Perey and F. G. Perey \cite{PEREY19761} for neutron and proton, and from  J. R. Huizenga and G.Igo \cite{HUIZENGA1962462} for the $\alpha$ particles. The angular distributions of $\alpha$ production due to compound nucleus evaporation reaction are given in Fig. \ref{fig:inclspec} and energy spectra at different angles for 25.7 MeV and 19.7 MeV are given in Figs. \ref{fig:enespec26} and \ref{fig:enespec20}, respectively which are represented by solid lines. The calculated angular distributions and energy spectra well above grazing angles match with the experimental data at 25.7, 22.7 and 19.7 MeV energies where the contribution is only from compound nuclear evaporation process.
\subsection{\label{sec:level40}Continuum Discretized Coupled Channel Calculations}
The non-capture or elastic breakup cross-sections are calculated with CDCC
using the code FRESCO version 3.1 \cite{fresco,sfresco}. $^6$Li nucleus was assumed as a two body $\alpha$ + $\textit{d}$ cluster. The continuum above the breakup threshold of $^6$Li $\longrightarrow$ $\alpha$ + \textit{d} were discretized into momentum bins of width $\Delta k$ = 0.1 fm$^{-1}$. The continuum momentum bins were truncated at $\epsilon_{max}$ = 9.25 MeV. Each continuum or resonance state was further binned into 40 equal k-bins. The relative orbital angular momentum L = 0, 1, 2 and 3 were included in the calculations. In addition, the 1$^+$, 2$^+$ and 3$^+$ resonances for L = 2 with experimental widths at 2.186 MeV, 4.312 MeV and 5.65 MeV respectively, were also included. The binding potential for the $\alpha$ and $\textit{d}$ clusters were taken from Ref. \cite{hirata}. The resonance potentials were included from Ref. \cite{tores03}. The cluster folding potentials for $\alpha$ + $^{51}$V and $\textit{d}$ + $^{51}$V at respective energies as per mass (2/3 of $E_{lab}$ for $\alpha$ and 1/3 of $E_{lab}$ for deuteron) were generated from Ref. \cite{haixia06,avrigeanu10}. The breakup cross-sections at different energies are given in Table \ref{tab:incl} and angular distributions are plotted in Fig. \ref{fig:inclusive}.

\subsection{\label{sec:level42}Coupled Reaction Channel Calculations}
Transfer and inelastic reactions are studied to explore the internal properties of nuclei, the arrangement of nucleons somewhere inside the nucleus. Therefore, for this purpose, transfer reactions are used to calculate single particle structure of nuclei and extraction of spectroscopic factors. In this section, we estimate the contribution of 1\textit{n}, 1\textit{p} and 1\textit{d} direct transfer cross-section for the reactions $^{51}$V($^{6}$Li, $^{5}$Li)$^{52}$V,  $^{51}$V($^{6}$Li, $^{5}$He)$^{52}$Cr and $^{51}$V($^{6}$Li, $^{4}$He)$^{53}$Cr, respectively. The parent and daughter nucleus incorporates coupling from different possible bound and excited states. To perform the coupled channel calculations, we need structural information of participating nuclei. In addition to this, other important physical input parameters are (1) optical potentials for incoming and outgoing distorted waves; (2) a peculiar identification of the overlap functions which typically represent single particle states in a Wood-Saxon potential with an intention to reproduce a bound state by re-adjusting depth of the binding potential; (3) spectroscopic factors corresponding to probability of finding core state within the composite state.

An incoming wave has both elastic and inelastic components. The wave function can be represented as 
 \begin{equation}
 \psi = \phi_{\alpha}(r)\chi_{\alpha}(R) + \phi_{\alpha'}(r)\chi_{\alpha'}(R)
 \end{equation} 
 Here $\phi_{\alpha}$(R) and $ \phi_{\alpha'}$(r) are the ground and excited state wave functions of projectile. The $\alpha$  represents incoming partition with projectile $\textit{a}$ and target \textit{A}. The functions $\chi_{\alpha}$(R) and  $\phi_{\alpha'}$(r) represent relative motion between projectile and target in numerous internal states. Of course the total wave function $\psi$ satisfies the Schrodinger equation (E-H)$\psi$ = 0. A set of two equations is obtained by projecting this equation onto different internal states:
 \begin{equation}\label{eq:cceq2}
\begin{cases}
 (E-{\epsilon_\alpha-K_\alpha-U_{\alpha\alpha}}) \chi_\alpha(R) = U_{\alpha\alpha'}\chi_{\alpha'}(R) \\

  (E-{\epsilon_\alpha'-K_\alpha'-U_{\alpha'\alpha'}}) \chi_\alpha'(R) = U_{\alpha'\alpha}\chi_{\alpha}(R)
\end{cases}
 \end{equation}
where U$_{\alpha\alpha}$ and U$_{\alpha\alpha'}$ are the coupling potentials. In a typical approach for calculation of coupled reaction channel (CRC), these two equations in Eq. \ref{eq:cceq2} are solved `exactly' to obtain $\chi_{\alpha}$(R) and $\chi_{\alpha'}$(R). 
In many cases, the inelastic component of the wave function is weakly coupled to ground state and thus this virtue opens the door for the approximated solution for the above equation. This can be obtained by making its inelastic component part to zero i.e.
  \begin{equation} 
  (E-{\epsilon_\alpha-K_\alpha-U_{\alpha\alpha}})\chi_\alpha(R) \approx 0
  \end{equation}
The resulting function $\chi_\alpha(R)$ is then inserted into the second equation to calculate $\chi_{\alpha'}(R)$. This is called 1-step Distorted Wave Born Approximation (DWBA) calculation. Further, this calculated $\chi_{\alpha'}(R)$ can again be inserted into 1$^{st}$ part of Eq. \ref{eq:cceq2} to have an iterative solution for $\chi_\alpha(R)$. In all the upcoming theoretical calculations by \textsc{FRESCO} we have used iterative method for full CRC calculations till the absolute difference between successive s-matrix elements becomes less than 0.01\%. The coupling to few discrete channels or continuum channels depends on the excitation energy for the particular channel. The direct deuteron transfer to lower discrete states is insignificant due to higher positive \textit{Q} value thus continuum coupling is important. Other transfer channels have contribution of discrete as well as continuum channels. The optical potentials for the incoming channel were obtained from elastic scattering data \cite{kumawat20} and is same for all transfer channels. 
\begin{figure}[h]
\includegraphics[scale=0.40]{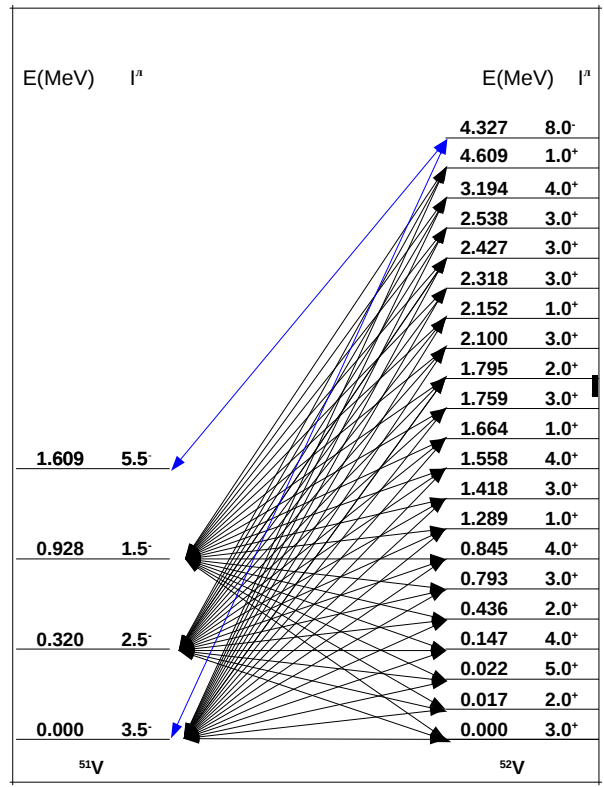}
\caption{\label{fig:coupling}Schematic diagram of coupling of different levels used in calculation for 1\textit{n} transfer.}
\end{figure}
\subsubsection{\label{sec:level43}  1n Transfer}
The 1\textit{n} transfer followed by breakup of $^{6}$Li $\Rightarrow$ $^{5}$Li(\textit{p} + $\alpha$) + \textit{n} can contribute to $\alpha$ production. The CRC calculations of 1\textit{n} transfer resulted in considerable contribution to inclusive $\alpha$ cross section by the reaction $^{51}$V($^{6}$Li, $^{5}$Li)$^{52}$V. The Woods–Saxon form factors were used with reduced radii r$_0$ = 1.25 fm and diffuseness \textit{a} = 0.65 fm for projectile as well as target bound state potentials. The spin-orbit interaction was included with standard depth of 6 MeV. The depth of the real potential was allowed to vary to reproduce experimental neutron binding energies.  The finite range transfer approximation in \textsc{FRESCO} \cite{fresco} was used for the calculations in post form.  Full complex remnant term were used with two way coupling scheme. The other important parameter,  spectroscopic factor for the projectile  was taken from Ref. \cite{cohen67}. The spectroscopic factors for target were incorporated from the work of O. Karban $\textit{et. al.}$ \cite{karban87}. The calculations were performed for \textit{n} transfer to the 2$p_{3/2}$, 1$f_{5/2}$, 2$p_{1/2}$ and 1$g_{9/2}$ orbits of available model space of $^{52}$V with the assumption of closure of 1$f_{7/2}$ subshell in the ground state. The continuum coupling above neutron bound state was considered with angular momentum of L = 0$\rightarrow$5 $\hbar$ and using equal linear momentum bins up to $\approx$ 15 MeV of energies.
The schematic picture of target overlaps and discrete states for coupling are shown in Fig. \ref{fig:coupling} which were part of the calculations. The contribution of 1\textit{n} transfer is mentioned in Table \ref{tab:incl}. The total angular distribution for 1\textit{n} transfer having sum of all participating states including continuum states in calculation, is presented in Fig. \ref{fig:inclusive}. 
\begin{figure}[h]
\includegraphics[scale=0.42]{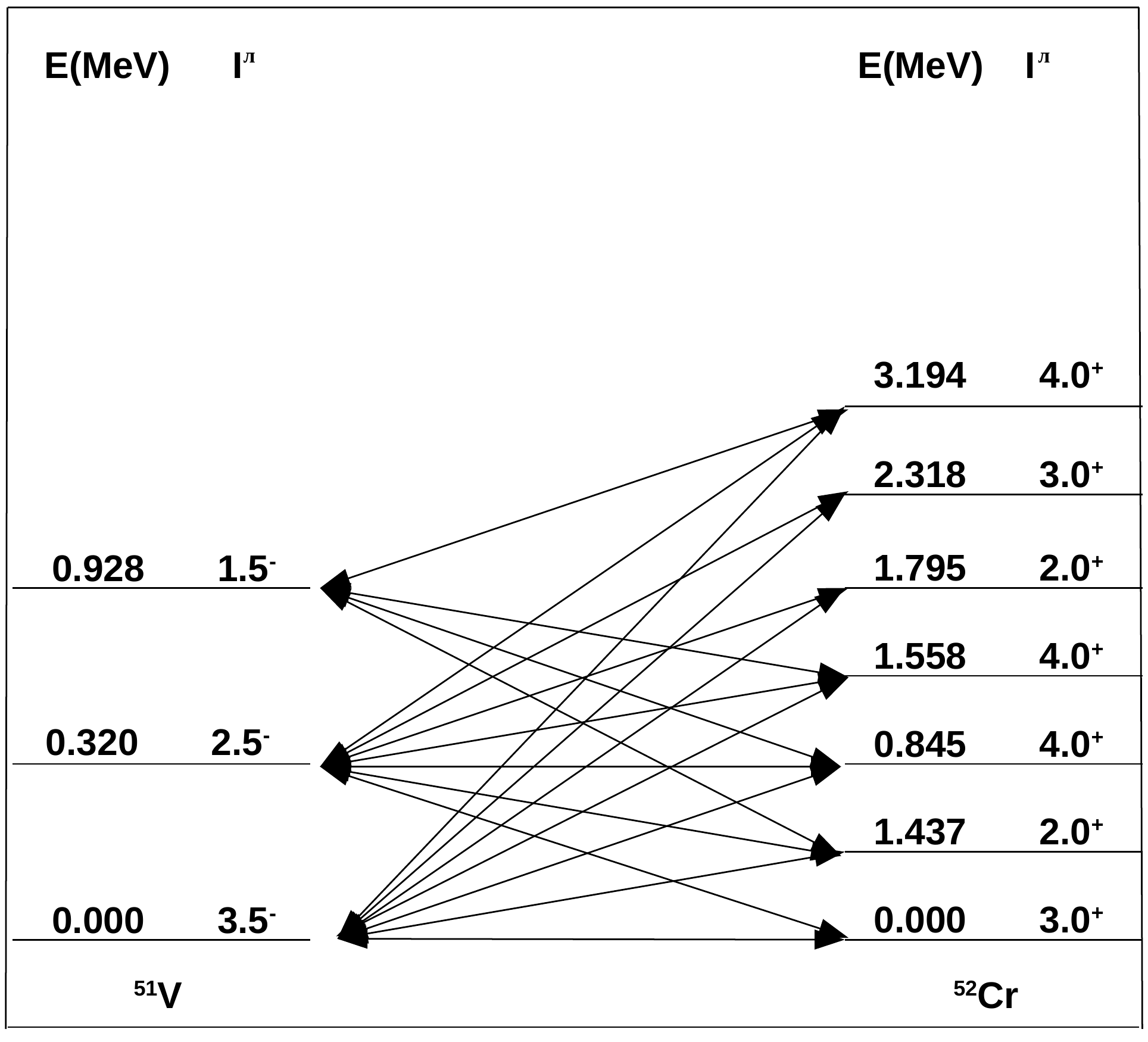}
\caption{\label{fig:couplingp}Schematic diagram of coupling of different levels used in calculation for 1\textit{p} transfer.}
\end{figure}

\subsubsection{\label{sec:level6}  1p Transfer}
\begin{figure}[h]
\includegraphics[scale=0.38]{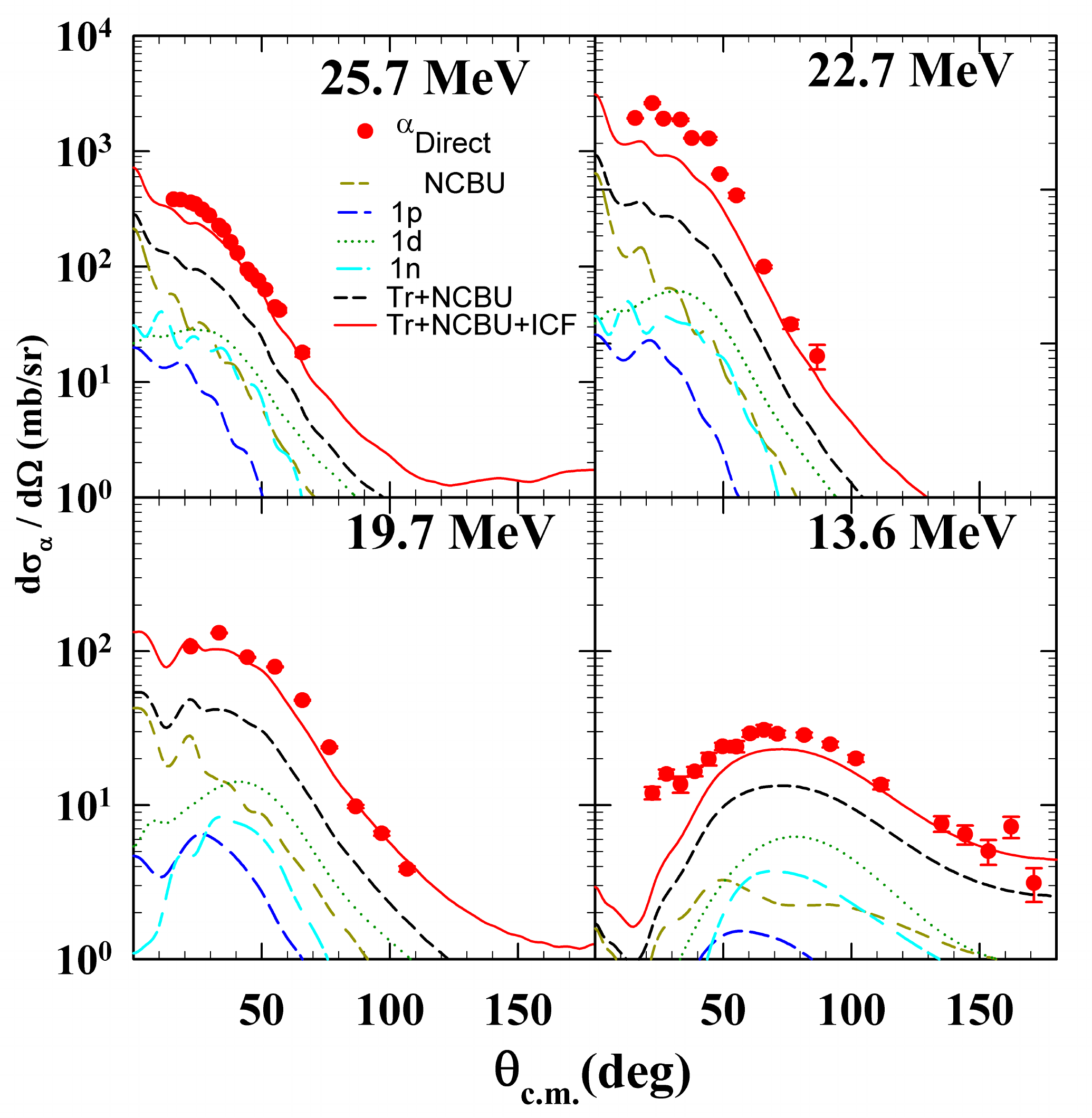}
\caption{\label{fig:inclusive}The contribution from different possible reaction channels contributing to direct $\alpha$ production i.e. non-capture breakup (dark yellow short-long dashed line), 1\textit{p} transfer (blue dashed line), 1\textit{n} transfer (cyan long-short dashed line), 1\textit{d} transfer (green dotted line), Transfer + NCBU (black short dashed line), and ICF + Transfer + NCBU is represented by solid line. Experimental inclusive direct $\alpha$ cross-section data are represented with red solid circles. ICF-$\alpha$ is estimated as described in Section \ref{sec:level4}.}
\end{figure}

In order to study the 1\textit{p} stripping or transfer contribution through $^{51}$V($^{6}$Li,$^{5}$He)$^{52}$Cr reaction, CRC calculations have been performed. The calculations were performed for \textit{p} transfer to the 1$f_{7/2}$, 2$p_{3/2}$ orbits in the model space of $^{52}$Cr. The schematic picture of target overlaps and states for coupling is shown in Fig. \ref{fig:couplingp}. The spectroscopic factors for target were taken from Ref. \cite{pellegrini73}. The spectroscopic factors were taken as one to get the contribution of higher states up to 9.5 MeV. The continuum couplings upto 12 MeV above bound state was included which give rise to small contribution. The 1\textit{p} transfer cross-section is mentioned in Table \ref{tab:incl}. The total angular distribution for 1\textit{p} stripping having sum of all participating states is presented in Fig. \ref{fig:inclusive}.
\subsubsection{\label{sec:level7} d-Cluster Transfer}
The CRC calculations were performed for 1\textit{d} transfer contribution which leads to direct $\alpha$ production through $^{51}$V($^{6}$Li, $^{4}$He)$^{53}$Cr reaction. The spectroscopic factors for the projectile and target were taken as one. As the \textit{Q} value for this channel is quite high (14.745 MeV), the contribution of transfer to low lying discrete states was negligible. Thus, the transfer to continuum coupling of $^{53}$Cr states up to $\approx$ 12 MeV energies above deuteron binding energy with equal momentum bins of bin width $\Delta$\textit{k} = 0.1 fm$^{-1}$ and angular momentum of L = 0 to 7 $\hbar$ were considered. The cross-section contribution of \textit{d} transfer is reported in Table \ref{tab:incl}. The angular distribution for 1\textit{d} transfer with sum of all continuum states in calculation is presented in Fig. \ref{fig:inclusive}.

The angular distributions in Fig. \ref{fig:inclusive} show individual contributions from NCBU and transfer channels and their addition is represented by Tr + NCBU. The ICF $\alpha$ contributions were added by multiplying the Tr + NCBU spectra with a factor of ($\sigma_{\alpha}^{cal.}$/($\sigma_{\alpha}^{cal.}$-$\sigma_{\alpha}^{ICF}$) where cross-sections are given in Table \ref{tab:incl} and $\sigma_{\alpha}^{ICF}$ calculations are described in Section \ref{sec:level4}. An overall agreement with experimental data is good and small difference might be due to uncertainties in spectroscopic factors and inclusion of limited number of states in transfer calculations.
\section{\label{sec:level8} Systematic Study  }
\begin{figure}[H]
\includegraphics[scale=0.4]{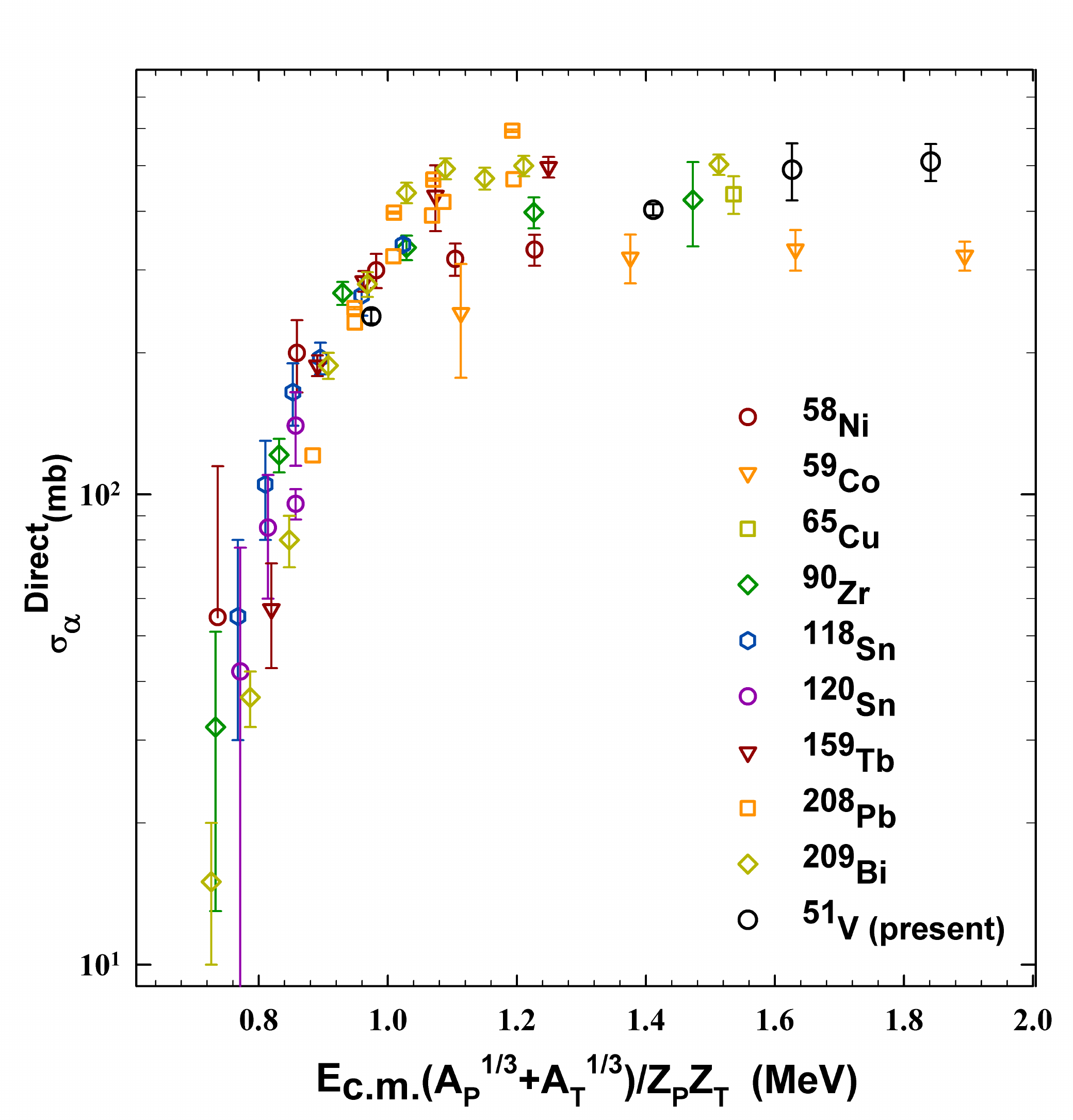}
\caption{\label{fig:alpha_syst} Experimental direct $\alpha$ particle cross-section data for $^{6}$Li projectile with different targets. The cross-sections for various targets are taken from literature $^{58}$Ni,$^{118}$Sn,$^{120}$Sn \cite{PFEIFFER1973545}, $^{59}$Co \cite{souza09},$^{65}$Cu \cite{SHRIVASTAVA2006463}, $^{90}$Zr \cite{kumawat10}, $^{159}$Tb \cite{PhysRevC.88.064603}, $^{208}$Pb \cite{signorini,PhysRevC.67.044607}, $^{209}$Bi \cite{PhysRevC.85.014612}.  The present work for $^{6}$Li + $^{51}$V is represented with black hollow circles.}
\end{figure}

 The direct $\alpha$ cross sections for $^{6}$Li projectile with various targets have been compared in Fig. \ref{fig:alpha_syst} along with present work. A reduction approach to obliterate the effect of the Coulomb barrier for different targets at varying energies \cite{gomes05} was adopted. The energy is reduced to scale $E_{c.m.}(A^{1/3}_P$+$A^{1/3}_T$)/Z$_P$Z$_T$, where \textit{P} and \textit{T} stand for projectile and target respectively, \textit{Z} and \textit{A} are charge and mass of involved nuclei. The direct contributions presented here show an overall universal trend and saturation above barrier energies with a noticeable difference for different targets. As the contribution to direct $\alpha$ particles is from many transfer channels; these have an influence of nuclear structure. The reduced cross-sections do not show suppression for light and mid mass targets compared with heavy mass targets.
\begin{figure}[H]
\includegraphics[scale=0.6]{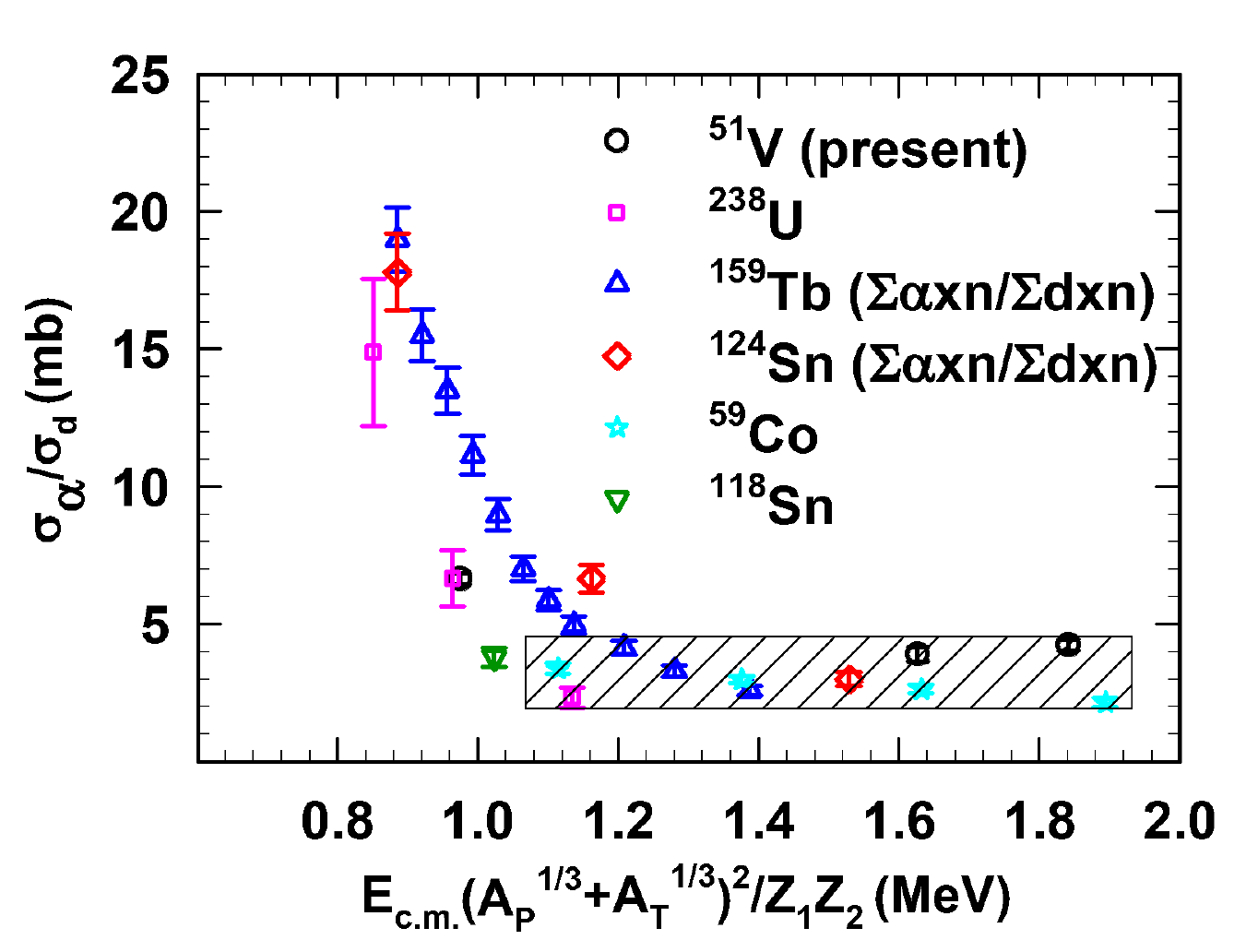}
\caption{\label{fig:ratio} A ratio of $\alpha$ production cross-section to deuteron production cross-section for different targets around Coulomb barrier. Experimental data are taken from Ref. $^{6}$Li + $^{238}$U \cite{PhysRevC.99.024620}, $^{6}$Li + $^{124}$Sn \cite{PhysRevC.98.014601}, $^{6}$Li + $^{159}$Tb \cite{PhysRevC.88.064603}, $^{6}$Li + $^{59}$Co \cite{souza09}, $^{6}$Li + $^{118}$Sn \cite{PFEIFFER1973545}.}
\end{figure}
 A systematic study of $\alpha$  and deuteron production cross-sections was carried out and ratio of these cross-sections is plotted in Fig. \ref{fig:ratio}. The cross-section for $^{51}$V (present study), $^{59}$Co \cite{souza09}, $^{118}$Sn \cite{PFEIFFER1973545} and $^{238}$U \cite{PhysRevC.99.024620} are from direct observation of these particles but for $^{124}$Sn \cite{PhysRevC.98.014601} and $^{159}$Tb \cite{PhysRevC.88.064603} the data are taken from ICF residue cross-sections. The ratio suggests many more reaction channels for $\alpha$ production (breakup, \textit{n}, \textit{p} transfer followed by breakup, \textit{d} transfer and breakup fusion etc.) compared to deuteron production (breakup, $\alpha$ transfer, and breakup fusion etc.). We have to also consider the effect of lower deuteron binding leading to its split into \textit{p} + \textit{n} particles.
\section{\label{sec:level9} Summary }
In summary, inclusive $\alpha$ production cross-sections have been measured for $^{6}$Li + $^{51}$V system around Coulomb barrier over a wide angular range and large $\alpha$ particle yields were observed. Statistical model calculations have been performed to separate the compound nuclear contributions in the $\alpha$ particle spectra and to get the direct $\alpha$ contributions. Non-capture break-up calculations using CDCC method, CRC calculations for \textit{n}, \textit{p} and \textit{d} transfers and ICF $\alpha$ estimations using fusion cross-sections were performed to disentangle the contribution from these channels to $\alpha$ production. Addition of all these channel cross-sections leading to $\alpha$ production reproduce angular and energy distributions, and integral cross-sections reasonably well. Deuteron cluster transfer gives negligible contribution to discrete states due to high \textit{Q} value and transfer to continuum gives significant contribution. Transfer to continuum enhances \textit{n} and \textit{p} transfer along with discrete state contributions, marginally. ICF $\alpha$ cross-sections were deduced from fusion cross-sections. The total calculated direct $\alpha$ cross-sections are in good agreement with the experimental data. The kinematic disentanglement of $\alpha$ particles suggest that breakup fusion is dominant over transfer as the positive \textit{Q} value boosted $\alpha$ particles are not significant. The dominance of breakup process and breakup fusion contributions over the direct transfer process in the production of $\alpha$ particles could be perhaps related to $^{6}$Li having a smaller binding energy when compared to that of $^{7}$Li. More experiments of particle-$\gamma$ coincidence are required with suitable targets where dis-entanglement is irrefutable. Direct $\alpha$ particle cross-sections show a universal trend on average with a noticeable difference with different targets possibly due to structure effects in transfer channels contributing to $\alpha$ production but no suppression was observed in reduced direct $\alpha$ cross-section for light mass targets compared to heavy mass targets. Ratio of $\alpha$  and deuteron production cross-sections for various targets shows much higher cross-section for $\alpha$ production close to barrier compared to deuteron production due to difference in Coulomb barrier and more reaction channels for $\alpha$ production. The ratio saturates above barrier energies perhaps due to saturation of number of open channels. 
\begin{acknowledgments}
The authors thank to BARC/TIFR Pelletron staff for excellent delivery of the beam and the support of the Department of Atomic Energy, Government of India, under project No. 12P-R\&D-TFR-5.02-0300. One of the author (CJ) would like to acknowledge Department of Science and Technology (DST), Government of India for awarding her Inspire Fellowship. The authors V.V.P. and S.K. acknowledge the financial support from Young Scientist Research grant and Senior Scientist program, respectively, from the Indian National Science Academy (INSA), Government of India, in carrying out these investigations.
\end{acknowledgments}
%  \bibliography{alpha}
%  \end{document}

\end{document}